\documentclass[conference]{IEEEtran}
\IEEEoverridecommandlockouts
\usepackage{cite}
\usepackage{amsmath,amssymb,amsfonts}
\usepackage{algorithmic}
\usepackage{graphicx}
\usepackage{textcomp}
\usepackage{xcolor}
\usepackage{authblk}
\usepackage[version=4]{mhchem}
\usepackage{url}
\usepackage{hyperref}
\hypersetup{hidelinks,
colorlinks=true,urlcolor=blue}
\def\BibTeX{{\rm B\kern-.05em{\sc i\kern-.025em b}\kern-.08em
    T\kern-.1667em\lower.7ex\hbox{E}\kern-.125emX}}
\begin{document}

\title{FlavorDB2: An Updated Database of Flavor Molecules}

\author[1]{Nishant Grover\textsuperscript{$\dagger$}}
\author[1]{Mansi Goel\textsuperscript{$\dagger$}}
\author[2]{Devansh Batra}
\author[3]{Neelansh Garg}
\author[4]{\\Rudraksh Tuwani}
\author[1]{Apuroop Sethupathy}
\author[*]{Ganesh Bagler\textsuperscript{1}}
\affil[1]{\textit{\ Center for Computational Biology, Indraprastha Institute of Information Technology (IIIT-Delhi), New Delhi, India}}
\affil[2]{\textit{\ Department of Information Technology, Netaji Subhas University of Technology, New Delhi, India}}
\affil[3]{\textit{\ Department of Computer Science, University of South California, USA}}
\affil[4]{\textit{\ Department of Biostatistics, Harvard T.H. Chan School of Public Health, Boston University, USA}}
\affil[*]{\textit{\ Corresponding author: Ganesh Bagler, bagler@iiitd.ac.in}}

\maketitle

\begin{abstract}

Flavor is expressed through interaction of molecules via gustatory and olfactory mechanisms. Knowing the utility of flavor molecules in food and fragrances, it is valuable to add a comprehensive repository of flavor compounds characterizing their flavor profile, chemical properties, regulatory status, consumption statistics, taste/aroma threshold values, reported uses in food categories, and synthesis. FlavorDB2 (https://cosylab.iiitd.edu.in/flavordb2/) is an updated database of flavor molecules with an user-friendly interface. This repository simplifies the search for flavor molecules, their attributes and offers a range of applications including food pairing. FlavorDB2 serves as a standard repository of flavor compounds.
\end{abstract}


\section{Introduction}
 Flavor compounds form the physical basis for a wide range of gustatory and olfactory sensations experienced by humans \cite{fisher2007food}. By interacting with the biological machinery, these compounds evoke a complex perception of taste and odor. Flavor perception is an emergent property of the complex biological system, the complete understanding of which still eludes us \cite{fisher2007food,shepherd2015neuroenology,malnic1999combinatorial,mouritsen2015science,newcomb2013genetics}. While simple associations between molecular properties and perception indicate the chemical basis of flavor \cite{fisher2007food}, such knowledge is mainly heuristics-based and remains unstructured. 
\\

Creating a detailed catalog of various aspects of flavor compounds offers an approach leading to a systemic interpretation of flavor sensation. Such a repository, on one side, would have the molecular descriptors enumerating physico-chemical properties of the flavor compounds. On the other side, it would have natural sources (in which these compounds are found), sensory features (such as flavor percepts), mechanisms of synthesis, regulatory status, and applications (such as foods categories in which the compound is utilized), among others. Such a compilation opens up space for asking a variety of data-driven questions. FlavorDB was created to incorporate multidimensional aspects of flavor molecules and represent their molecular features, flavor profiles, and details of natural sources \cite{garg2018flavordb}. 
\\

\begin{figure}[!htb]
\includegraphics[width=0.5\textwidth]{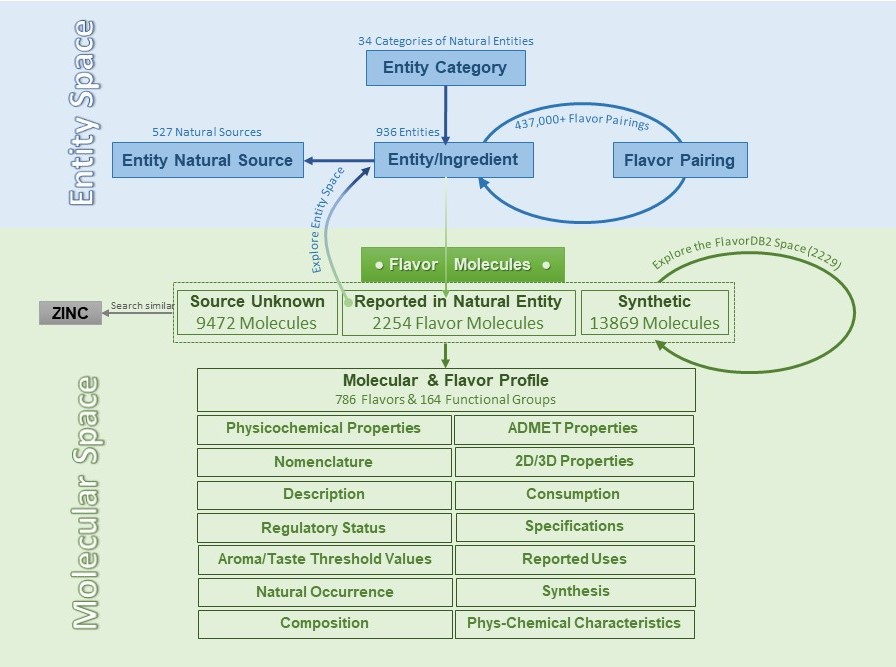}
\caption{FlavorDB2 provides significantly advanced attributes of flavor compounds and search mechanisms. It presents a comprehensive repository of flavor compounds through a user-friendly interface and interlinked search engines for exploring the flavor universe.}
\label{flavordb2}
\end{figure}

Despite being the most comprehensive repository of flavor compounds, FlavorDB missed out some of the critical details of flavor compounds available in the literature. FlavorDB2 expands the flavor space, thereby providing a significant advancement from FlavorDB. FlavorDB2 is a combination of ‘entity space’ and ‘molecular space’ (fig. \ref{flavordb2}) where former provides the entities utilized in food from natural sources and latter shows the molecular and flavor profiles. One of the databases with closely aligned objectives, FooDB(http://foodb.ca), collates molecules from food ingredients. Some of the resources that have attempted to create a data compilation on specific aspects of flavor include Flavornet (Arn and Acree, 1998), BitterDB \cite{wiener2012bitterdb}, SuperSweet \cite{ahmed2010supersweet}, SuperScent \cite{dunkel2009superscent}. Among the other efforts of data compilation are those targeted at nutritional factors (NutriChem), polyphenols (Phenol-Explorer), and the medicinal value of food \cite{scalbert2011databases, rothwell2013phenol, jensen2015nutrichem, neveu2010phenol}. The uniqueness of FlavorDB lies in its extensive coverage of flavor compounds and nutritional information to more accurately convey ingredient’s flavor profile. 
\\

FlavorDB2 collates information from various sources, including FooDB, Flavornet, SuperSweet, BitterDB, and Fenaroli's Handbook of Flavor Compounds to build a comprehensive and structured repository of flavor molecules. While the number of compounds (25,595), ingredients (936), and ingredient categories (34) remain the same as that in FlavorDB, FlavorDB2 adds depth by adding a wide range of molecular features.

\section{database overview}
The details of expanded content to enrich the characterization of flavor molecules are provided below. We have exhaustively accounted for an array of features of flavor compounds making it as comprehensive as possible through this expansion.  

\subsection{Collection and Compilation of New Data}
\begin{figure*}[!htb]
\includegraphics[width=\textwidth]{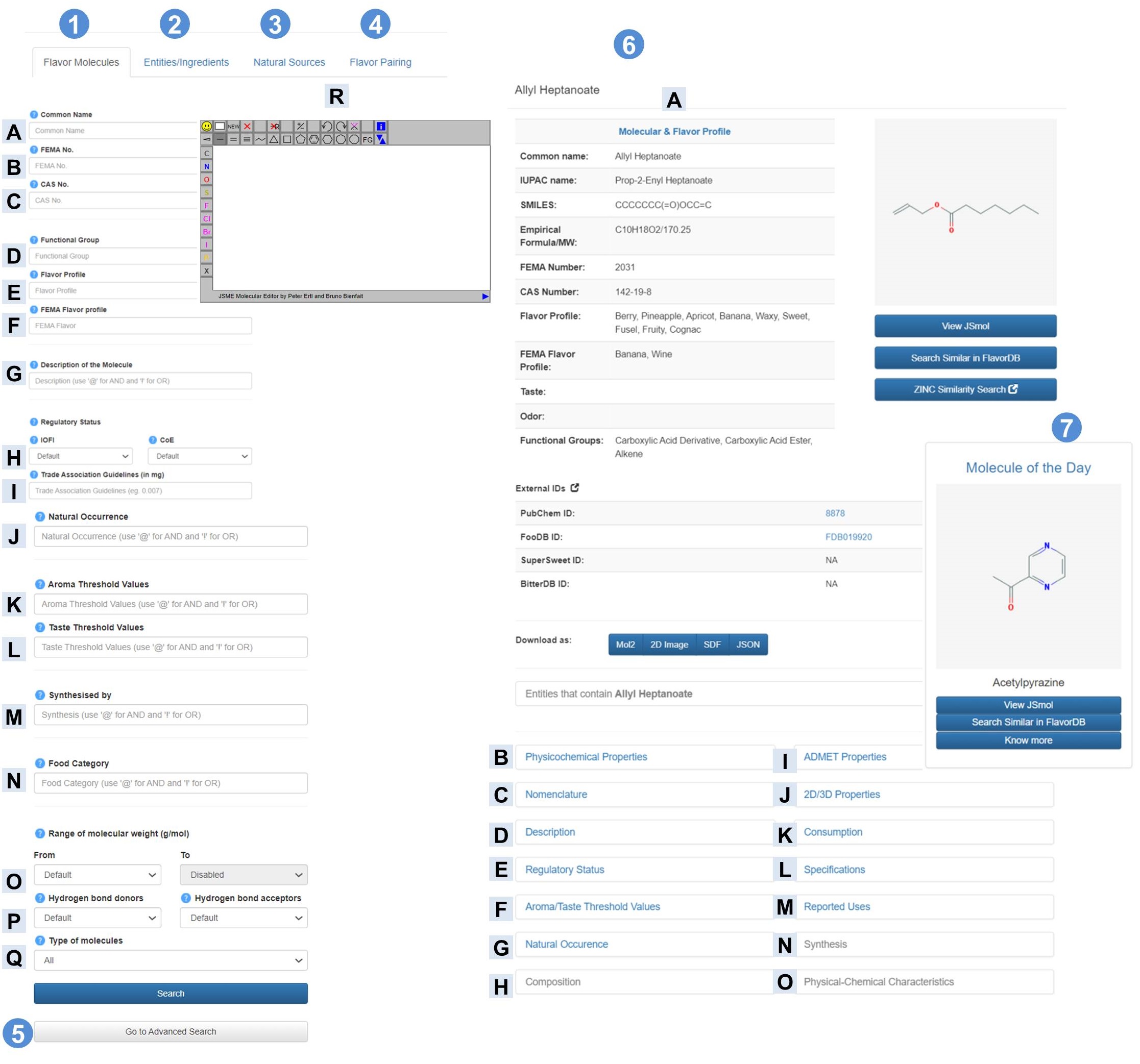}
\caption{Schematic view of FlavorDB2 highlighting the search and molecular properties of the data. (1) Flavor Molecules Search (A. Common name, B. FEMA Number, C. CAS Number, D. Functional Group, E. Flavor Profile, F. FEMA Flavor Profile, G. Description of the Molecule, H. Regulatory Status, I. Trade Association Guidelines, J. Natural Occurrence, K. Aroma Threshold Values, L. Taste Threshold Values, M. Synthesized by, N. Food Category, O. Range of Molecular Weight, P. Hydrogen bond donors and acceptors, Q. Type of molecules, R. JSME Molecular Editor), (2) Entities/Ingredients Search, (3) Natural Sources Search, (4) Flavor Pairing Search, (5) Advanced Search, (6) Molecular Profile and Chemical Properties (A. Molecular and Flavor Profile, B. Physicochemical Properties, C. Nomenclature, D. Description, E. Regulatory Status, F. Aroma/Taste Threshold values, G. Natural Occurrence, H. Composition, I. ADMET Properties, J. 2D/3D Properties, K. Consumption, L. Specifications, M. Reported Uses, N. Synthesis, O. Physical-Chemical Characteristics), (7) Molecule of the Day.}
\label{schema}
\end{figure*}

We sought to compile a detailed profile of each of the naturally occurring synthetic compound comprising of a wide variety of molecular attributes: (i) variety of nomenclatures/IDs, (ii) flavor description, (iii) regulatory status, (iv) taste/aroma threshold values, (v) natural occurrence, (vi) composition, (vii) consumption statistics, (viii) specifications, (ix) reported uses in food categories, and (x) synthesis (see fig. \ref{schema}). These features represent vital attributes of flavor compounds that bring value to the users of the database. Please refer the Figures S1, S2, S3, S4, S5, and S6 in Supplementary data for the statistics of flavor consumption, food category, regulatory status, and their chemical properties.

\subsection{Molecular Nomenclature/IDs}
One of the issues in dealing with flavor compounds is variation in their nomenclature. A flavor compound could be identified with any of the following nomenclatures/IDs: common name, IUPAC (International Union of Pure and Applied Chemistry), SMILES (Simplified Molecular-Input Line-Entry System), Empirical Formula, CAS (Chemical Abstracts Service) Number, FEMA (Flavor Extract Manufacturers Association) Number, FL (FLAVIS) Number, NAS (National Academy of Sciences) Number, COE (Council of Europe) Number, EINECS (European Inventory of Existing Commercial Substances) Number, or JFCFA (Joint FAO/WHO Expert Committee on Food Additives) Number. Given this redundancy in naming, one of the challenges in the flavor molecular space is to arrive at the right compound given one of its diverse IDs. Hence, we have implemented search for all major molecular IDs used for flavor compounds (in basic and/or advanced search). The nomenclature field also provides a list of synonyms helping further in reducing the ambiguity in the identification of the desired compound.

\subsection{Description}
Beyond a list of flavor descriptors previously made available in FlavorDB, the ‘Description’ field provides a brief nuanced description of the flavor. For example, for compound allyl hexanoate, reported to be found in pineapple among other ingredients, the description field reads as – ‘A colorless liquid with sweet, pineapple-like taste, and fruit-like aroma’. 

\subsection{Regulatory Status}
This field provides regulatory/safety details about compounds as per the regulatory bodies such as COE, JECFA, FDA (Food and Drug Administration), IOFI (International Organization of the Flavor Industry), and Trade Association Guidelines. The information includes acceptable quantity/concentration of the compound (in parts per million (ppm), parts per billion (ppb), or mg) and nature identical status, etc. For allyl hexanoate, apart from providing the acceptable level of usage (COE-5ppm; FDA-21; CFR-172.515; JECFA-0-0.13 mg/kg; Trade Association Guidelines-10.749 mg), the field mentions that the compound is of natural origin.

\subsection{Aroma/Taste Threshold Values}
This field provides the information of aroma/taste threshold values at specific concentrations with nuanced details of the taste/aroma. For allyl hexanoate, this field reads as ‘Taste attributes at 10 ppm; sweet, juicy, fresh, pineapple, and fruity’.

\subsection{Natural Occurrence}
The natural occurrences field provides details of the natural ingredients/sources in which the compound is reported to be found. For allyl hexanoate, this field reads as ‘Developed in a baked potato, mushroom, and pineapple’.

\subsection{Consumption}
This field provides the compound’s consumption details in terms of annual (in pounds) and individual consumption (mg/kg/day). The reported annual consumption is 16033.3 lb for allyl hexanoate, whereas the individual consumption is 0.01358 mg/kg/day.

\subsection{Specifications}
This field provides details of the compound such as appearance, solubility, acid value, specific gravity, refractive index, assay, boiling, and melting points. For allyl hexanoate, this field suggests that its appearance is ‘colorless to light-yellow’; it is soluble in ethanol; its specific gravity is 0.884-0.890 (25°C), and boiling point is 185°C.

\subsection{Reported Uses}
A compound is used in various food categories (such as Alcoholic Beverages, Hard Candy, Chewing gum, etc.) as a flavoring agent. This field provides the compound’s various usage as per FEMA. Allyl hexanoate is reported to be used in alcoholic beverages, gravies, baked goods, hard candy, chewing gum, meat products, frozen dairy, non-alcoholic beverages, gelatins puddings, and soft candy.

\subsection{Synthesis}
A compound is synthesized through a chemical process, either from a natural source or via synthetic mechanisms. This field provides details of the source compound and the process used for its synthesis. Allyl hexanoate is reported to be synthesized by esterification of \textit{n}-caproic acid with allyl alcohol in the presence of concentrated \ce{H2SO4} or naphthalene-$\beta$-sulfonic acid in benzene under a nitrogen blanket.

\section{use cases}
\subsection{Searching Flavor Compounds by Alternative IDs}
FlavorDB2 can be searched for flavor compounds using a wide variety of molecular IDs/nomenclatures namely common name, IUPAC, SMILES, CAS Number, FEMA Number, FL Number, NAS Number, EINECS Number, and JFCFA Number. In the presence of using diverse molecular numbering systems, this enables reaching the desired flavor compounds in the repository. 

\subsection{Flavor Compounds Used in Food Category}
One may search the ‘food category’ to know all compounds associated to a category. For a compound against each category, the usual and maximum concentration at which it is used is also listed. Users can create a compounded query to get the flavor compounds by adding multiple food categories using @ (AND) or ! (OR) operator. 

\subsection{Flavor Compounds based on Regulatory Status}
Flavor compounds can be searched based on their IOFI values namely natural, artificial, nature identical, not nature identical, artificial and identical. One can also search based on COE values namely approved, approved in some quantity, used provisionally, unknown. FlavorDB2 will generate the list of flavor molecules based on the input query. 

\subsection{Flavor Compound Synthesis} 
To identify a flavor compound that involves a given chemical in its synthesis process, one could search by the name(s) of the chemical(s). For example, searching by 'acetic acid' opens a list of flavor compounds involving acetic acid in their synthesis. One can create a compounded query to add more compounds with the help of @ (AND) or ! (OR) operator. The query can be performed not only on compounds but also on chemical processes (such as ‘esterification’, ‘distillation’) or similar terms used in the synthesis field.  

\section{implementation}
Consistent with FlavorDB, FlavorDB2 database has been designed using MySQL (https://www.mysql.com). Django (https://www.djangoproject.com), a Python web development framework, has been used for webserver development \cite{holovaty2009definitive}. To enhance the functionality of FlavorDB2, jQuery, Bootstrap, D3.js, and Google Charts libraries along with HTML, CSS and Javascript were used. Apache HTTP Server route requests to Django and helps to enable data compression for faster page load times. FlavorDB2 can be viewed in the latest versions of Google Chrome, Firefox, Opera, Internet Explorer, and Microsoft Edge.

\section{conclusion}
Investigation of molecules and components associated with flavor sensation is of keen interest for food pairing \cite{blumenthal2008big} and aroma blending \cite{chambers2013associations}. FlavorDB2 provides the detailed information of flavor molecules using ‘entity space’ and ‘molecular space’ which will help in molecular gastronomy, predicting odor from chemical properties, and food pairing applications \cite{keller2017predicting, spence2013touch, this2002molecular}.

In spite of our earnest attempts, FlavorDB2 is an updated and advanced repository of flavor molecules in which several new attributes are added to analyze the flavor compounds from various perspectives. The data of flavor molecules of an ingredient is limited due to the unavailability of the information. These advanced properties simplify the search for flavor molecules. 

\section{availability}
The FlavorDB2 database is freely accessible at \url{https://cosylab.iiitd.edu.in/flavordb2/}. Users are not required to register or log in to access any feature available in the database. 

\section*{Acknowledgment}
G.B. thanks Indraprastha Institute of Information Technology (IIIT-Delhi) for providing computational facilities and support. G.B. thanks Technology Innovation Hub (TiH) Anubhuti for the research grant. N.G., D.B., N.G., R.T. and A.S. were Summer Research Interns and M.G. is a Research Scholar in Dr. Bagler’s lab (Complex Systems Laboratory) at Center for Computational Biology. All research interns are thankful to IIIT-Delhi for the support. M.G. thanks IIIT-Delhi for the fellowship. This research was partially supported by the Infosys Centre for Artificial Intelligence.   

\bibliographystyle{IEEEtran}
\bibliography{flavordb2}

\end{document}